\documentclass[cup5b]{caps}
\usepackage{graphicx}
\usepackage{amssymb}
\usepackage{ociwsymp4}   

\HeadText{P. E. Nissen}

\newcommand{\kmprs}  {\mbox{\rm km\,s$^{-1}$}}
\newcommand{\feh} {\mbox{\rm [Fe/H]}}
\newcommand{\ofeh} {\mbox{\rm [O/Fe]}}
\newcommand{\alphafeh} {\mbox{\rm [$\alpha$/Fe]}}
\newcommand{\OFe} {\mbox{\rm [O/Fe]}}
\newcommand{\NaFe} {\mbox{\rm [Na/Fe]}}
\newcommand{\MgFe} {\mbox{\rm [Mg/Fe]}}
\newcommand{\SiFe} {\mbox{\rm [Si/Fe]}}
\newcommand{\CaFe} {\mbox{\rm [Ca/Fe]}}
\newcommand{\TiFe} {\mbox{\rm [Ti/Fe]}}
\newcommand{\NiFe} {\mbox{\rm [Ni/Fe]}}
\newcommand{\MnFe} {\mbox{\rm [Mn/Fe]}}
\newcommand{\ZnFe} {\mbox{\rm [Zn/Fe]}}

\newcommand{\EuBa} {\mbox{\rm [Eu/Ba]}}
\newcommand{\teff}  {\mbox{$T_{\rm eff}$}}
\newcommand{\logteff}  {\mbox{log$T_{\rm eff}$}}

\newcommand{\forOI} {\mbox{\rm [O\,{\sc i}]}}

\newcommand{\Mgone} {Mg\,{\sc i}}
\newcommand{\Sione} {Si\,{\sc i}}

\newcommand{\Caone} {Ca\,{\sc i}}
\newcommand{\Tione} {Ti\,{\sc i}}

\newcommand{\Mnone} {Mn\,{\sc i}}
\newcommand{\Feone} {Fe\,{\sc i}}
\newcommand{\Fetwo} {Fe\,{\sc ii}}

\newcommand{\Znone} {Zn\,{\sc i}}

\newcommand{\Batwo} {Ba\,{\sc ii}}
\newcommand{\Eutwo} {Eu\,{\sc ii}}

\newcommand{\Vrot}   {\mbox{$V_{\rm rot}$}}
\newcommand{\Vtot}   {\mbox{$V_{\rm tot}$}}

\newcommand{\sigU}  {\mbox{$\sigma_{U}$}}
\newcommand{\sigV}  {\mbox{$\sigma_{V}$}}
\newcommand{\sigW}  {\mbox{$\sigma_{W}$}}
\newcommand{\Vlag}  {\mbox{$V_{\rm lag}$}}
\newcommand{\Mbol}  {\mbox{$M_{\rm bol}$}}

\begin{document}

\pagenumbering{arabic}

\author[]{POUL E. NISSEN\\Department of Physics and Astronomy, University of
Aarhus, Denmark}

\chapter{Thin and Thick Galactic Disks}

\begin{abstract}

Studies of elemental abundances in stars belonging to the thin and the
thick disk of our Galaxy are reviewed. Edvardsson et al.
(1993) found strong evidence of \alphafeh\ variations among
F and G main sequence stars 
with the same \feh\ and interpreted these differences as due to radial gradients in
the star formation rate in the Galactic disk. Several recent studies suggest,
however, that the differences are mainly due to a separation in \alphafeh\
between thin and thick disk stars, indicating that 
these populations are discrete
Galactic components, as also found from several kinematical studies.
Further evidence of a chemical separation between the thick and the thin
disk is obtained from studies of \MnFe\ and the ratio between $r$- and $s$-process
elements. The interpretation of these new data in terms of formation scenarios
and time scales for the disk and halo components of our Galaxy is discussed.

\end{abstract}

\section{Introduction}

A long-standing problem in studies of Galactic structure and evolution
has been the possible existence of a population of stars with 
kinematics, ages, and chemical abundances
in between the characteristic values for the halo
and the disk populations. Already at the Vatican Conference on
Stellar Populations (O'Connell 1958), an {\it intermediate Population II}
was introduced as stars with a velocity component perpendicular
to the Galactic plane on the order of $W \approx 30$~\kmprs .
Using the $m_1$ index of F-type stars,  Str\"{o}mgren (1966) later 
defined intermediate Population II as stars having
metallicities in the range $-0.8 < \feh < -0.4$, and from a
discussion of the extensive $uvby$-$\beta$ photometry
of Olsen (1983), he concluded that the
intermediate Population II consisted of old, 10--15 Gyr stars
with velocity dispersions (\sigU , \sigV , \sigW ) significantly
higher than those of the younger, more metal-rich disk stars
(Str\"{o}mgren 1987, Table 2).

In a seminal paper, Gilmore \& Reid (1983) showed that the distribution 
of stars in the direction of the Galactic South Pole could not be fitted
by a single exponential, but required at least two disk components ---
a {\it thin disk} with a scale height of 300 pc and a {\it thick disk} with
a scale height of about 1300 pc. They furthermore identified
intermediate Population II with the sum of the metal-poor end of the
old thin disk and the thick disk.  Following this work, it has
been intensively discussed if the thin and thick disks
are discrete components of our Galaxy or if there is a more
continuous sequence of stellar populations connecting the
Galactic halo and the thin disk. For a comprehensive review
and a discussion of possible formation scenarios, the reader is referred
to Majewski (1993). 

Quite a strong indication of the thin and thick Galactic disks as discrete 
populations with respect to kinematics and age came from the detailed
abundance
survey of Edvardsson et al. (1993). On the basis of the large $uvby$-$\beta$
catalogs of Olsen (1983, 1988), main sequence stars in the temperature
range $5600\,{\rm K}\, < \teff < 7000$\,K were selected and divided into 9 metallicity
groups ranging from $\feh \approx -1.0$ to $\sim +0.3$. In each metallicity
group the $\sim 20$ brightest stars were observed. Hence, there is no kinematical
bias in the selection of the stars. As shown by Edvardsson et al. 
(1993, Fig.\,16b) and as first discussed by Freeman (1991), there is an
abrupt increase in the $W$ velocity dispersion of the stars when an age
of 10 Gyr is passed. The same was found by Quillen \& Garnett (2001),
who reanalyzed the Edvardsson et al. sample 
using space velocities based on Hipparcos data (ESA 1997) and ages from
Ng \& Bertelli (1998). As seen from their Figure\,2, the velocity dispersions
are fairly constant for ages between 3 and 9 Gyr:
(\sigU , \sigV , \sigW ) $\simeq$ (35, 23, 18)\,\kmprs , 
corresponding to the thin disk,
whereas for ages between 10 and 15 Gyr the dispersions
are (\sigU , \sigV , \sigW ) $\simeq$ (60, 50, 40)\,\kmprs ,
where the velocity dispersion $\sigW = 40$\,\kmprs\  corresponds quite well
to the scale height of the Gilmore \& Reid thick disk.
About the same values were obtained by Nissen (1995) on the
basis of the original Edvardsson et al. (1993) data.
Furthermore, he derived rotational lags with respect to the
local standard of rest (LSR), $\Vlag \simeq -10$\,\kmprs\ for the thin disk
and $\Vlag \simeq -50$\,\kmprs\ for the thick disk. 

Although the Edvardsson et al. data for the kinematics of stars
in the solar neighborhood belonging to
the thin and thick Galactic disks
refer to 189 stars only, the values derived agree quite 
well with other recent investigations.  For example, Soubiran, Bienaym\'{e}, 
\& Siebert (2003) derive (\sigU , \sigV , \sigW ) =
($63 \pm 6, 39 \pm 4, 39 \pm 4$)\,\kmprs and a rotational lag 
$\Vlag = -51 \pm 5$\,\kmprs for the thick disk based on Tycho-2
proper motions (H\o g et al. 2000) and ELODIE (Baranne et al. 1996)
spectra for a sample of 400 stars in directions toward
the Galactic North Pole.

In the following, we review recent studies of the chemical composition
of Galactic disk stars. As we shall see, there is increasing evidence
that the thin and thick disks overlap in metallicity in the
range $-0.8 < \feh < -0.4$ but are separated in 
\alphafeh , where $\alpha$ refers to the $\alpha$-capture elements.
Furthermore, recent studies suggest that the two disk components
are also separated in \MnFe\
and \EuBa , i.e. the $r$- to $s$-process ratio.
Hence, the chemical studies support the interpretation of the
thin and thick disks as discrete components of our galaxy 
formed at separated epochs and having different evolution time scales.

\section{The {\Large $\alpha$-}Capture Elements}
It is well known that $\alpha$-capture elements like O, Mg, Si, and Ca
are overabundant by a factor of 2 to 3 relative to Fe in the large 
majority of metal-poor halo stars, i.e \alphafeh = +0.3 to +0.5.
In the disk \alphafeh\
decreases with increasing \feh\ to zero at solar metallicity, an
effect that is normally explained in terms of delayed production
of iron by Type Ia supernovae (SNe) in the disk phase. As the release of  
Type Ia products occurs with a time delay of typically 1 Gyr, the
metallicity at which \alphafeh\ starts to decline depends critically
on the star formation rate. Hence, \alphafeh\ may be used as 
``a chemical clock'' to date the star formation process in the Galaxy.

\subsection{The Edvardsson et al. Survey}
\begin{figure*}
\centering
\includegraphics[width=0.90\columnwidth,angle=0]{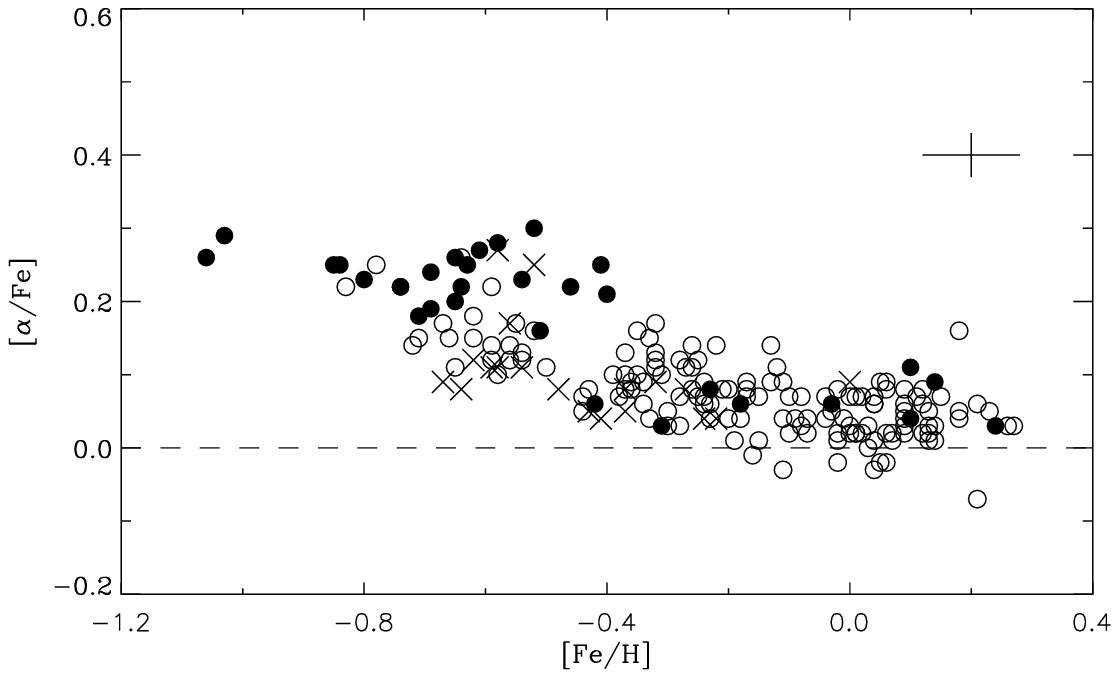}
\vskip 0pt \caption{
\alphafeh\ vs. \feh\ for the Edvardsson et al. (1993) stars.
\alphafeh\ is defined as  $\frac{1}{4} (\MgFe + \SiFe + \CaFe + \TiFe )$.
Stars shown with filled circles have a mean galactocentric distance in their 
orbits $R_m < 7$~kpc. Open circles refer to stars with 
$7\,{\rm kpc}\, < R_m < 9$~kpc, and crosses refer to stars with $R_m > 9$~kpc.
Typical 1 $\sigma$ error bars, referring to differential abundances
at a given \feh , are indicated.}
\label{nis.fig1}
\end{figure*}

The Edvardsson et al. (1993) survey provided clear evidence for a
scatter in \alphafeh\ among disk stars with the same \feh . This is
shown in Figure\,\ref{nis.fig1},
where \alphafeh\ is plotted as a function of \feh .
\alphafeh\ is defined as the average abundance of Mg, Si, Ca, and Ti
with respect to Fe, and was measured with a differential
precision of about 0.03 dex for stars having about the same metallicity.
Such a high precision can be obtained when the selected stars 
belong to relatively narrow ranges in \teff\ and gravity,
like the Edvardsson et al. sample, and when the abundance ratios are
derived from weak absorption lines having about the same 
dependence of \teff\ and gravity,
such as the \Mgone , \Sione , \Caone , \Tione\ and \Feone\ lines. 
In other cases, like \ofeh, where the oxygen abundance is derived 
from the \forOI\ $\lambda$6300 line, the derived abundance ratio
is more sensitive to errors in atmospheric parameters and the
structure of the model atmospheres (e.g. 3D effects; Nissen et al. 2002). 
This is why oxygen abundances were not
included when calculating the average $\alpha$-element abundance.
It should also be emphasized that 
the absolute abundances and the overall
trend of \alphafeh\ with \feh\ may be affected by non-LTE effects.

As seen from Figure\,\ref{nis.fig1},
\alphafeh\ for stars in the metallicity
range $-0.8 < \feh < -0.4$ is correlated with the mean 
galactocentric distance $R_m$ in the stellar orbit.
Stars with $R_m > 9$~kpc tend to have lower \alphafeh\ than stars
with $R_m < 7$~kpc, and stars belonging to the solar circle lie
in between. Assuming that $R_m$  is a statistical measure of the
distance from the Galactic center at which the star was born,
Edvardsson et al. explained the \alphafeh\ variations 
as due to a  star formation rate that declines with  
galactocentric distance. In other words, Type Ia SNe
start contributing with iron at a higher \feh\ in the inner parts 
of the Galaxy than in the outer parts. As we shall see in the
following, the \alphafeh\ variations may, however, also be
interpreted in terms of systematic differences between
thin and thick disk stars.

\subsection{Recent Studies of {\large $\alpha$-}Capture Elements}
Gratton et al. (1996) were the first to point out that the
variations in \alphafeh\ could be interpreted in terms of
systematic differences between the chemical composition of
thin and thick disk stars. Later, Gratton et al.
(2000) studied these differences in more detail;
equivalent width data from Zhao \& Magain (1990),
Tomkin et al. (1992), Nissen \& Edvardsson (1992), and
Edvardsson et al. (1993) were reanalyzed in a homogeneous 
way and used to derive Fe/O and Fe/Mg ratios. When the stars
are plotted in a [Fe/O] vs. [O/H] diagram, two groups of
disk stars with [O/H]$> -0.5$  appear:
thin disk stars with [Fe/O]$> -0.25$
and thick disk stars with [Fe/O]$< -0.25$. The two groups
show a large degree of overlap in [O/H]. Gratton et al.
interpreted this as evidence for a sudden decrease in
star formation rate during the transition between the thick and
thin disk phases, allowing Type Ia SNe to enrich the interstellar
gas with Fe without any increase in O and Mg  due to the
absence of Type II SNe.

An even more clear chemical separation between thick and thin 
disk stars has been obtained by Fuhrmann (1998, 2000).
For a sample of nearby stars with $5300\,{\rm K}\, < \teff < 6600$\,K
and $3.7 < \log\,g < 4.6$, he derived Mg abundances from \Mgone\
lines and Fe abundances from \Feone\ and \Fetwo\ lines.
In a \MgFe\ vs. \feh\ diagram, stars with thick disk kinematics
have $\MgFe \simeq +0.4$ and \feh\
between $-1.0$ and $-0.3$. The thin disk stars show a well-defined 
sequence from $\feh \simeq -0.6$ to +0.4
with \MgFe\ decreasing from +0.2 to 0.0. Hence, there is a
clear \MgFe\ separation between thick and thin disk stars 
in the overlap region $-0.6 < \feh < -0.3$ with only a few
``transition'' stars. This is even more striking in a
diagram where [Fe/Mg] is plotted as a function of [Mg/H]
(Fuhrmann 2000, Fig.\,12). Fuhrmann's group of 16 thick disk stars
have total space velocities with respect to the LSR in the range
85 \kmprs\ $< \Vtot < 180$\,\kmprs\ and an average rotational 
lag of $\Vlag \simeq -80$\,\kmprs . It is unclear if
this low value is a selection effect. 

On the basis of stellar ages derived from evolutionary tracks in
\Mbol - \logteff\ diagrams, Bernkopf, Fiedler, \& Fuhrmann (2001) claim
that the maximum age of thin disk stars is about 9 Gyr, whereas the
thick disk stars have ages between 12 and 14 Gyr. These data 
agree well with their suggestion that the systematic difference
of \alphafeh\ is due to a hiatus in star formation 
between the thick and thin disk phases. However, Bernkopf et al. (2001)
derived ages for 7 stars only. Larger samples
of thick and thin disk stars should be dated before firm
conclusions regarding a hiatus in star formation can be drawn. 

Further evidence of a higher \alphafeh\ in thick disk stars
than in thin disk stars has been presented by Prochaska et al.
(2000), who made a thorough study of the chemical composition
of 10 G-type stars having $-1.2 < \feh < -0.4$ and maximum orbital
distances from the Galactic plane greater than 600\,pc.
Interestingly, \OFe , \SiFe, and \CaFe\ show a decline with
increasing \feh , which may be interpreted as a signature of 
enrichment of Type Ia SNe in the thick disk. This would mean
that the thick disk formed over a time scale $\geq 1$\,Gyr.
Prochaska et al. argue that such a formation time scale
would rule out most dissipational collapse scenarios for
the formation of the thick disk.

Recently, Feltzing, Bensby, \& Lundstr\"{o}m (2003b) have found strong evidence
for the presence of Type Ia SNe in the thick disk (see also papers at this 
meeting by Feltzing et al. 2003a and Bensby, Feltzing, \& Lundstr\"{o}m 2003). 
>From a sample of about 14\,000 dwarf stars in the solar neighborhood with 
metallicities and ages derived by Feltzing, Holmberg, \& Hurley (2001), they 
selected two samples with a high kinematical probability of belonging either 
to the thin or the thick disk. When plotted in a Toomre 
diagram\footnote{Sandage \& Fouts (1987) appear to be the first to
apply this type of diagram in a discussion of the escape velocity of the
Galaxy and to name it the ``Toomre energy diagram,'' recognizing that the
representation was due to A. Toomre (1980, private communication).} 
(Feltzing et al. 2003b, Fig.\,1) it is seen that the thin disk stars have 
total space velocities $\Vtot < +60$\,\kmprs , whereas the thick disk
stars are confined to the range 80 \kmprs\ $< \Vtot < 180$\,\kmprs .  

\begin{figure*}
\centering
\includegraphics[width=0.90\columnwidth,angle=0]{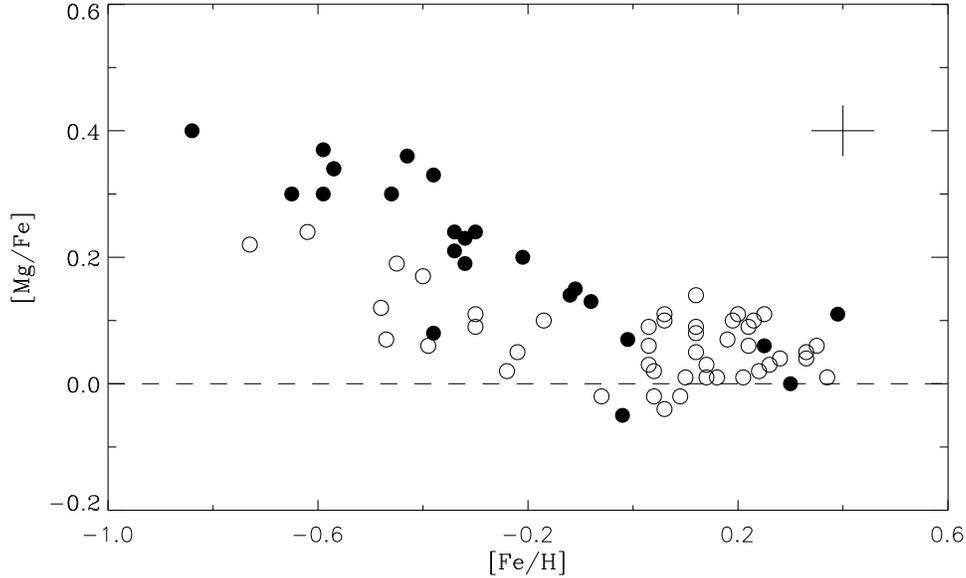}
\vskip 0pt \caption{
\MgFe\ vs. \feh\ from Feltzing et al. (2003b).
Stars shown with filled circles have thick disk kinematics;
open circles refer to thin disk stars.}
\label{nis.fig2}
\end{figure*}

Interestingly, the thick disk stars of Feltzing et al. (2003b)
are distributed over the whole metallicity range from $-1.0$
to $0.0$, and reach perhaps up to $\feh \simeq +0.4$. 
Below $\feh \simeq -0.4$, \alphafeh\ in the thick disk
stars is constant at a level of about 0.3 dex, and the 
thick disk is clearly separated from the thin disk in 
\alphafeh. Above $\feh = -0.4$, \alphafeh\ in the thick disk
declines and the two disks merge together. This is seen for
both Mg, Si, Ca, and Ti, but most clearly in \MgFe, as 
shown in Figure\,\ref{nis.fig2}.
Hence, star formation in the thick disk went
on long enough that Type Ia SNe started to enrich the gas 
out of which following generations of thick disk stars formed.

Attention is also drawn to a new work by Reddy et al. (2003).
A sample of 181 F--G dwarfs were selected
from the Olsen (1983, 1988) $uvby$-$\beta$ catalogs, and
the abundances of 27 elements were determined from
high-resolution spectra. Parallaxes and proper motions
were taken from the Hipparcos Catalogue (ESA 1997).
Nearly all stars studied have thin disk kinematics.
The $\alpha$-elements, O, Mg, Si, Ca, and Ti, show
\alphafeh\ to increase slightly with decreasing \feh\
in the range $-0.7 < \feh < 0.0$. When compared with
abundances for thick disk stars, mainly collected
from Fulbright (2000), the thick disk stars have 
\alphafeh\ about 0.15 dex higher than the thin disk stars
in the overlap region around $\feh \approx -0.5$. Hence,
the new work of Reddy et al. supports the thick-thin
$\alpha$-element separation discussed in this section.

\begin{figure*}
\centering
\includegraphics[width=0.90\columnwidth,angle=0]{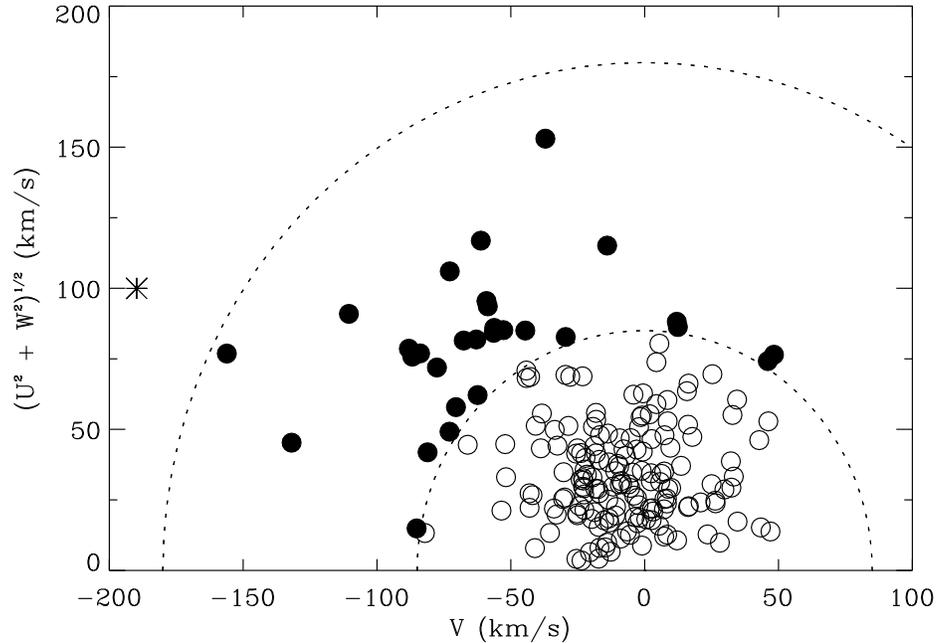}
\vskip 0pt \caption{
Toomre diagram for the Edvardsson et al. stars.
The two circles delineate constant total space velocities 
with respect to the LSR of
$\Vtot = 85$ and 180\,\kmprs , respectively, as used by
Fuhrmann (2000) to define a sample of thick disk stars.
According to this definition, filled circles are
thick disk stars, whereas open circles refer to thin disk stars.
One star, HD\,148816, shown by an asterisk, is classified
as a halo star.}
\label{nis.fig3}
\end{figure*}

\begin{figure*}
\centering
\includegraphics[width=0.90\columnwidth,angle=0]{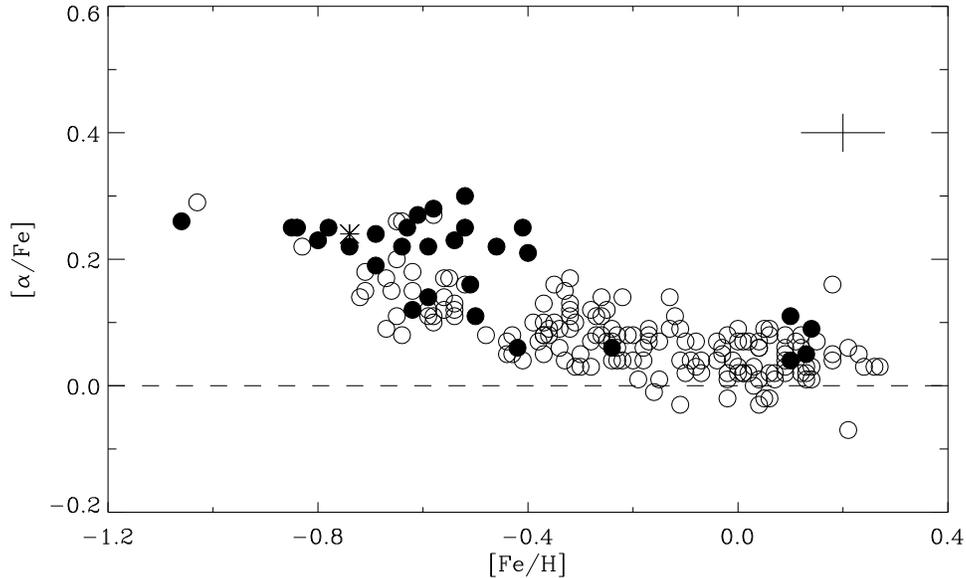}
\vskip 0pt \caption{
\alphafeh\ vs. \feh\ for the Edvardsson et al. (1993)
stars. As classified in Fig.\,3, filled circles are
thick disk stars and open circles are thin disk stars.
The  asterisk shows HD\,148816, the only
halo star in the sample.}
\label{nis.fig4}
\end{figure*}

In view of these new results on the separation of
\alphafeh\ between thin and thick stars, it is interesting 
to see if the scatter in \alphafeh\ for the 
Edvardsson et al. (1993) sample can be interpreted in
terms of thick-thin differences, instead of a correlation
with galactocentric distance (Fig.\,\ref{nis.fig1}). To investigate
this, I have plotted the Edvardsson et al. stars 
in a Toomre diagram (Fig.\,\ref{nis.fig3}) and divided them into
thin and thick disk stars according to the kinematical
definitions of Fuhrmann (2000). As seen from Figure\,\ref{nis.fig4},
much of the scatter can indeed be explained in terms
of thick-thin differences in \alphafeh . The separation
is not quite as clear as seen in the diagrams of
Fuhrmann (2000) and Feltzing et al. (2003b), but this
may be due to the fact that these authors have selected
kinematically well-separated groups of stars, whereas
the Edvardsson et al. sample is magnitude limited 
for a given metallicity bin and hence contains more
stars with kinematics in the thick-thin transition
region.

In a continuation of the work of Edvardsson et al.,
Chen et al. (2000) studied the chemical composition
of 90 F and G dwarf stars. They do not find any clear
\alphafeh\ separation between thin and thick disk stars.
As pointed out by Prochaska et al. (2000), this may,
however, be due to the fact that they selected
dwarf stars in the temperature range $5800\,{\rm K}\, < \teff < 6400$ K.
Hence, the old, more metal-rich thick disk stars with
$\teff < 5700$ K are not included. The few thick disk stars
in Chen et al. all have $\feh < -0.6$, i.e. they are
lying in a metallicity region where \alphafeh\ of the
thin disk merges with \alphafeh\ of the thick disk.   

\begin{figure*}
\centering
\includegraphics[width=0.90\columnwidth,angle=0]{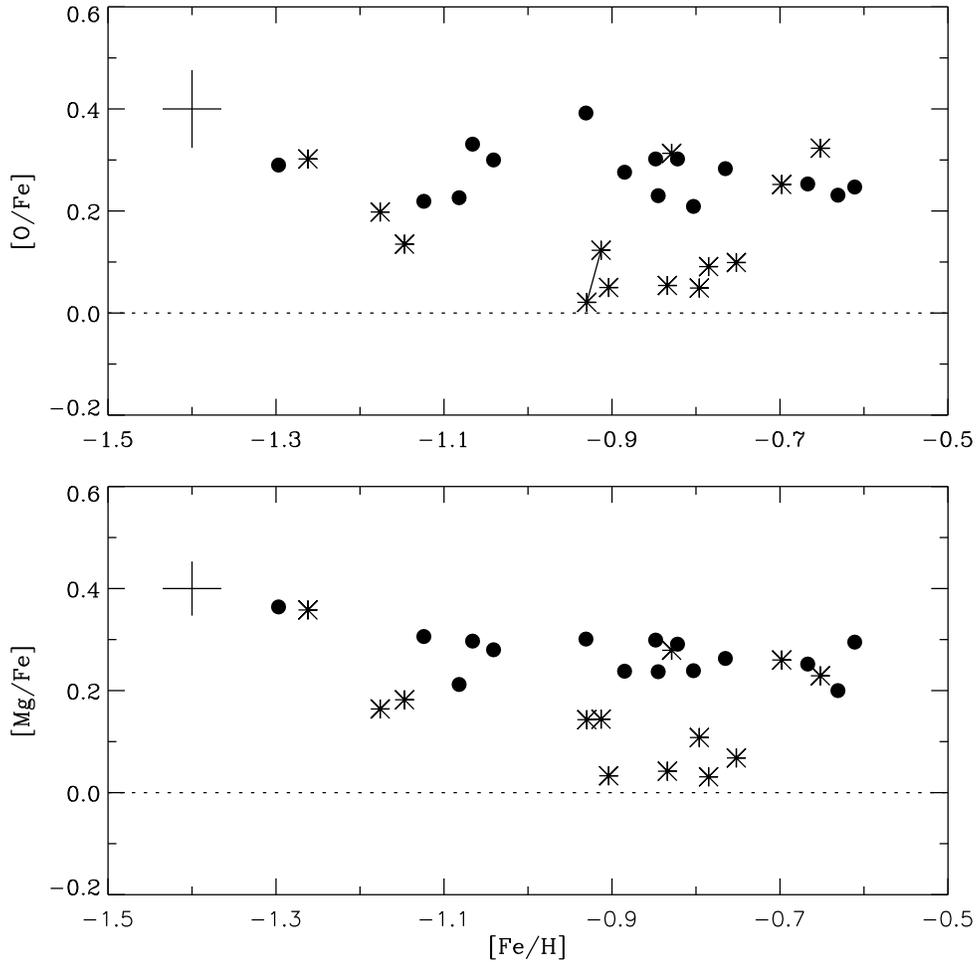}
\vskip 0pt \caption{
\OFe\ and \MgFe\ vs. \feh\ from Nissen \& Schuster (1997).
Filled circles refer to thick disk stars with a Galactic rotation
velocity component $\Vrot > 150$\,\kmprs , and asterisks refer to halo stars
with $\Vrot < 50$\,\kmprs . Two of the halo stars (connected
with a line) are components in a spectroscopic binary.}
\label{nis.fig5}
\end{figure*}

\subsection{A Comparison With \alphafeh\ in Halo Stars}
%\subsection{A Comparison With $\mathbf [\alpha/Fe]$ in Halo Stars}
Photometric and spectroscopic surveys of
high-velocity, main sequence and subgiant stars in the solar 
neighborhood by Nissen \& Schuster (1991), 
Schuster, Parrao, \& Contreras Mart\'{\i}nez (1993), and Carney et al. (1996)
have shown that the metallicity range $-1.5 < \feh < -0.5$ contains both
halo stars having a small velocity component in
the direction of Galactic rotation,
$\Vrot < 50$~\kmprs (where $\Vrot = V + 225$~\kmprs),
and thick disk stars with $\Vrot \simeq 175$~\kmprs .
Nissen \& Schuster (1997) selected such two groups of stars
with overlapping metallicities, and used
high-resolution, high signal-to-noise ratio spectra to determine 
abundance ratios of O, Na, Mg, Si, Ca, Ti, Cr, Fe, Ni, Y, and Ba
with a differential precision ranging from 0.02 to 0.07 dex
for 13 halo stars and 16 thick disk stars. Figure\,\ref{nis.fig5} shows the
results for \OFe\ and \MgFe\ vs. \feh . The same pattern is seen
for the other $\alpha$-elements, Si, Ca, and Ti, although with
a smaller amplitude for the abundance variations with respect
to Fe. As seen, all thick disk stars have a near-constant 
\alphafeh\ at a level of 0.3 dex, whereas the majority of
the halo stars have lower values of \alphafeh .

As discussed by Nissen \& Schuster (1997), there is a tendency
for the $\alpha$-poor halo stars to be on
larger Galactic orbits than halo stars with the same
abundance ratios as the thick disk stars.
>From this they suggest that the halo stars
with ``low-$\alpha$'' abundances have been formed in the outer part of
the halo or have been accreted from dwarf galaxies,
for which several models (Gilmore \& Wyse 1991; Tsujimoto et al. 1995;
Pagel \& Tautvai\v{s}ien\.{e} 1998) predict a solar $\alpha$/Fe ratio at
$\feh \simeq -1.0$ as a consequence of an early star formation burst
followed by a long dormant period. Recently, Shetrone et al. (2003)
and Venn et al. (2003) have
found that stars in dwarf spheroidal and irregular galaxies
with metallicities around $\feh \approx -1.0$ indeed have a
rather low \alphafeh\ ratio, supporting the view that the low-\alphafeh\ stars
belong to an accreted halo component.

Jehin et al. (1999) found two additional halo stars at $\feh \simeq -1.1$
having low values of \alphafeh . Among the more metal-poor
halo stars with $\feh < -1.4$, $\alpha$-poor stars are rare;
only a couple of cases have been found (Carney et al. 1997; 
King 1997). A more systematic study by Stephens \& Boesgaard (2002) of halo 
stars with unusual orbital properties, i.e. belonging to the ``outer'' or 
``high'' halo, did not reveal any new $\alpha$-poor 
stars, although a weak correlation between \alphafeh\ and
$R_{\rm apo}$ was detected. 

\begin{figure*}
\centering
\includegraphics[width=0.90\columnwidth,angle=0]{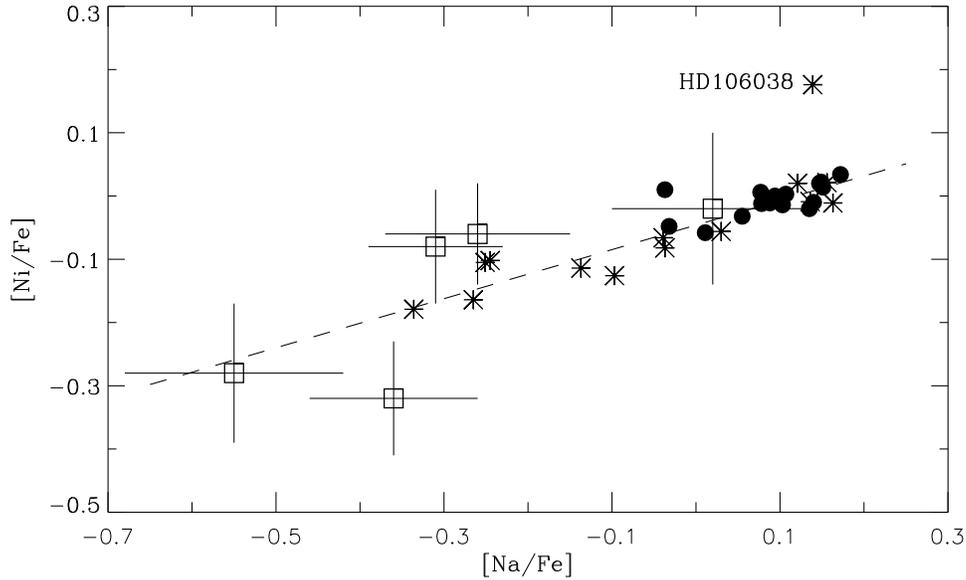}
\vskip 0pt \caption{
\NiFe\ vs. \NaFe .
Filled circles refer to to thick disk stars and asterisks to halo stars
with abundance ratios from Nissen \& Schuster (1997). The dashed
line is a fit to these data, excluding the peculiar star HD\,106038.
The squares with error bars are red giant stars in
dwarf spheroidal galaxies with abundances determined by Shetrone 
et al. (2003) and with metallicities in the range $-1.4 < \feh < -0.6$.} 
\label{nis.fig6}
\end{figure*}

Interestingly, the $\alpha$-poor stars are also deficient in
Na and Ni. Furthermore, there is a tight correlation
between \NiFe\ and \NaFe , as shown in Figure\,\ref{nis.fig6},
except for one peculiar Ni-rich halo star, HD\,106038, which 
is also very rich in Si and the $s$-process elements Y and Ba
(Nissen \& Schuster 1997). Furthermore, Figure\,\ref{nis.fig6} shows 
that dwarf spheroidal stars selected to have 
$-1.4 < \feh < -0.6$ tend to follow the relation delineated
by the $\alpha$-poor halo stars, hence supporting the idea that
the $\alpha$-poor stars are accreted from dwarf galaxies.
The reason for the correlation between
Na and Ni abundances is unclear, but it may be 
connected to the fact that the yields of both Na and the dominant
Ni isotope ($^{58}$Ni) depend upon the neutron 
excess (Thielemann, Hashimoto, \& Nomoto 1990).

\section{Manganese and Zinc}
\begin{figure*}
\centering
\includegraphics[width=0.90\columnwidth,angle=0]{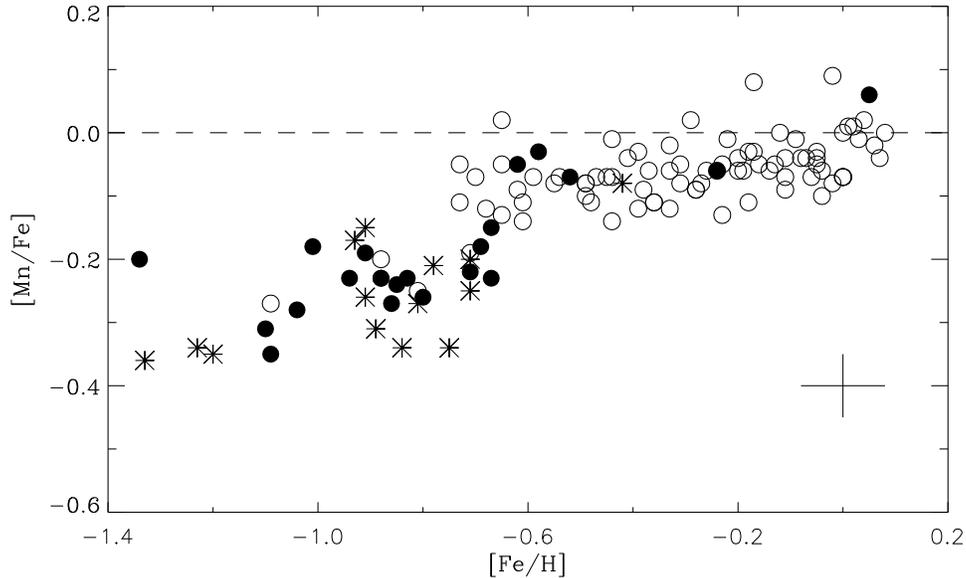}
\vskip 0pt \caption{
\MnFe\ vs. \feh\ with data from Nissen et al. (2000),
as corrected by Prochaska \& McWilliam (2000).  Open circles:
thin disk; filled circles: thick disk; asterisks: halo stars.}
\label{nis.fig7}
\end{figure*}

Among the iron-peak elements, Cr and Ni follow Fe very
closely (e.g., Chen et al. 2000), and there is no offset between
thick and thin disk stars (Prochaska et al. 2000). Manganese, on the
other hand, shows an interesting behavior. A detailed study
of the trend of \MnFe\ in disk and metal-rich halo stars was
published by Nissen et al. (2000) based on high-resolution
observations of the \Mnone\ $\lambda$6020 triplet. Nissen
et al., however, applied outdated data for the hyperfine structure 
of the \Mnone\ lines. Using modern hyperfine structure data,
Prochaska \& McWilliam (2000) found significant corrections
to the \MnFe\ values of Nissen et al. (2000). In Figure\,\ref{nis.fig7}
their revised data have been plotted with the same symbols
as in previous figures for halo, thin disk, and thick disk stars.
As seen, there is a steplike change in \MnFe\
at $\feh \simeq -0.6$. Thick disk stars with \feh\ below $-0.6$
have $\MnFe \simeq -0.3$, whereas thin disk stars with 
$-0.8 < \feh < -0.2$ have $\MnFe \simeq -0.1$. 
Due to the overlap in kinematics between the thick and the thin disk,
the few stars with $\feh > -0.6$ classified as thick disk
may, in fact, belong to the high-velocity tail of the thin disk.
The three stars with $\feh < -0.8$ classified as
thin disk have total space velocities with respect to the LSR
close to the 85\,\kmprs\ boundary that we have adopted as the separation
velocity between the thick and the thin disk; i.e. they may belong
to the thick disk. Altogether, the distribution of stars in
Figure\,\ref{nis.fig7} may be interpreted as a separation in \MnFe\ between the 
thin and the thick disk.

The trend of [Mn/Fe] is close to mirror that of
\alphafeh\  with respect to the [X/Fe] = 0 line
(compare Fig.\,\ref{nis.fig7} with Fig.\,\ref{nis.fig4}).
This suggests that Type Ia SNe is a main source for 
the production of Mn.
On the other hand, the eight $\alpha$-poor halo stars
from Nissen \& Schuster (1997), which are included in
Figure\,\ref{nis.fig7}, do not have higher \MnFe\ ratios
than the thick disk stars, which one would have expected
if Type Ia SNe were the main source of Mn production. 
Hence, it is not easy to understand the trend of \MnFe .
Probably, the underabundance of Mn is partly caused  
by a metallicity-dependent yield
due to a lower neutron excess in metal-poor stars
(Timmes, Woosley, \& Weaver 1995).

\begin{figure*}
\centering
\includegraphics[width=0.90\columnwidth,angle=0]{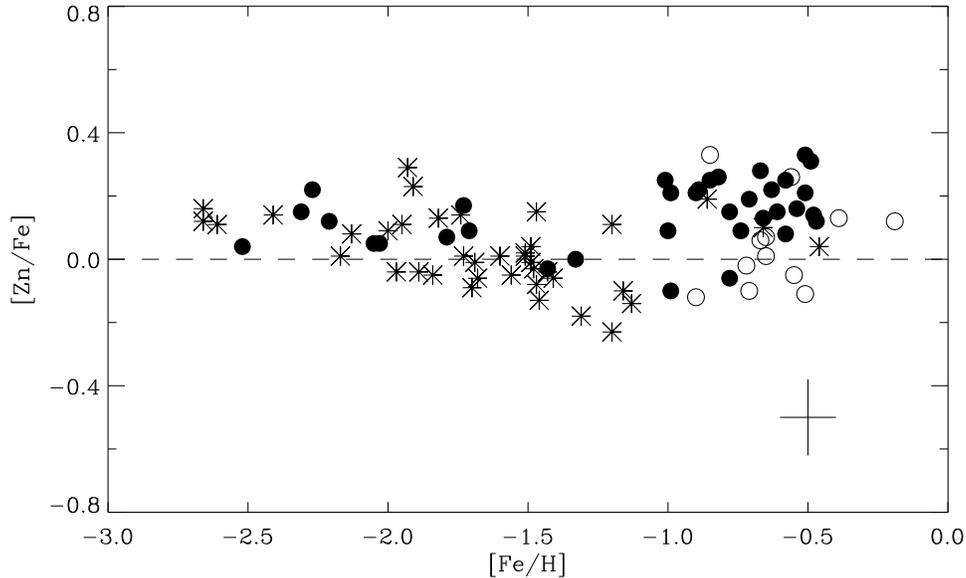}
\vskip 0pt \caption{
\ZnFe\ vs. \feh\ with data from Mishenina et al. (2002).
Open circles: thin disk; filled circles: thick disk; asterisks: halo stars.}
\label{nis.fig8}
\end{figure*}

Zinc is an interesting element with a number of possible
nucleosynthesis channels: neutron capture ($s$-processing)
in low- and intermediate-mass stars, as well as explosive
burning in Type II and Ia SNe (Matteucci et al. 1993).
Furthermore, zinc is a key element in studies of elemental
abundances of damped Ly$\alpha$ systems because
Zn is practically undepleted unto dust (e.g., Pettini et al. 1999).
In studies of damped Ly$\alpha$ systems, it is normally assumed that 
$\ZnFe \simeq 0.0$ 
in Galactic stars, as found by Sneden, Gratton, \& Crocker (1991) for 
the range $-3.0 < \feh < 0.0$,
although with quite a high scatter. Prochaska et al. (2000)
claim, however, that Zn is overabundant in thick
disk stars, $\ZnFe \simeq +0.1$. Recently, Mishenina et al. (2002)
have published a survey of Zn abundances in 90 disk and halo stars
based on equivalent widths of the \Znone\ $\lambda \lambda$4722.2, 4810.5,
6362.4 lines in high-resolution spectra of dwarf and giant stars.
Although the authors conclude that the data ``confirms the well-known
fact that the ratio \ZnFe\ is almost solar at all metallicities,''
there is in fact a hint of interesting structure in their \ZnFe\
trend. In Figure\,\ref{nis.fig8}, I have plotted the data of Mishenina et al. (2002)
using the total space velocity with respect to the LSR
to separate the stars into the halo, thin, and thick populations,
in the same way as in Figure\,\ref{nis.fig3}. As seen, there is a tendency
that thick disk stars in the metallicity range $-1.0 < \feh < -0.5$
are overabundant in Zn by as much as $\ZnFe \approx +0.2$. 
Furthermore, there may be a gradient in \ZnFe\ as a function of
\feh\ for the halo stars, with the highest \ZnFe\ for the most
metal-poor stars. Clearly, \ZnFe\ in halo and disk stars
should be further studied, 
if possible with smaller errors than those obtained by 
Mishenina et al. (2002).

An interesting detail from Figure\,\ref{nis.fig8} should be noted:
stars classified as thick disk occur down to metallicities
around $\feh \simeq -2.5$. Although it is difficult to
distinguish between thick disk and halo stars due to their
overlapping kinematics, it is interesting that studies of
large samples of metal-poor stars selected without kinematical
bias (Beers \& Sommer-Larsen 1995; Chiba \& Beers 2000) point
to the existence of thick disk stars at a rate of $\sim$10\%
relative to the halo population 
in the range $-2.2 < \feh  < -1.7$ and $\sim$30\%  for 
$-1.7 < \feh  < -1.0$.

\section{The {\Large $s$-} and {\Large $r$-}Process Elements}
\begin{figure*}
\centering
\includegraphics[width=0.90\columnwidth,angle=0]{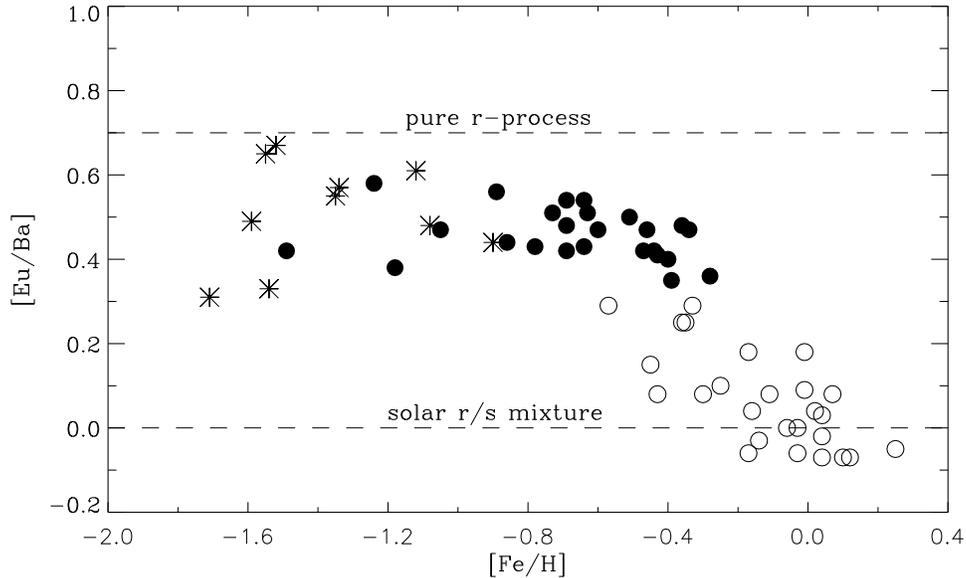}
\vskip 0pt \caption{
\EuBa\ vs. \feh\ with data from Mashonkina \& Gehren
(2000, 2001) and Mashonkina et al. (2003).
Open circles: thin disk; filled circles: thick disk; asterisks: halo stars.}
\label{nis.fig9}
\end{figure*}

A very interesting set of papers on barium and europium abundances
in cool dwarf stars have recently been published by 
Mashonkina \& Gehren (2000, 2001) and Mashonkina et al. (2003).
Their results are obtained from a non-LTE, differential
model atmosphere analysis of high-resolution, high signal-to-noise ratio spectra
of the \Batwo\ $\lambda \lambda$\,5853, 6496 lines and the
\Eutwo\ $\lambda$4129 line, taking into account hyperfine structure effects 
for the Eu line. Their data for the \EuBa\ ratio are plotted in 
Figure\,\ref{nis.fig9}. As seen, there is a rather clear separation  
between thick and thin disk stars. While thin
disk stars have a solar $r/s$ mixture at solar metallicity,
thick disk stars and some halo stars approach a pure 
$r$-process ratio. The steplike change in Eu/Ba 
around $\feh \approx -0.5$ from the thick to the thin disk
suggests a hiatus in star formation before the thin
disk developed, i.e. long enough to enable low-mass asymptotic giant branch 
(AGB) stars to produce Ba by the $s$-process. 

In addition to the separation in \EuBa\ between thick and thin
disk stars,
Mashonkina et al. (2003) claim that the slight decline in
\EuBa\ with increasing \feh\ for the thick disk stars
is significant. If real, this suggests a rather long time scale
(1.1 to 1.6\,Gyr) for the 
formation of the thick disk according to the chemical
evolution calculations of Travaglio et al. (1999). Finally,
a significant dispersion in \EuBa\ is seen for the halo stars,
which suggests a duration of the halo formation of about
1.5\,Gyr. Interestingly, Mashonkina et al. (2003) also find
a dispersion in \MgFe\ for the halo stars. At $\feh \approx -1.0$,
the average \MgFe\ in halo stars is lower than in 
thick disk stars, a result that agrees well with the findings
of Nissen \& Schuster (1997).

\section{Conclusions}
We have seen that disk stars in the metallicity range $-0.8 < \feh < -0.4$
have significant differences in the $\alpha$-element/Fe
abundance ratio, showing a variation of $\sim 0.2$\,dex in \MgFe\
and about 0.15\,dex in \SiFe , \CaFe, and \TiFe . These differences
were originally detected
by Edvardsson et al. (1993) and interpreted by them as due to a 
radial gradient in the star formation rate in the Galactic disk
causing the enrichment of iron-peak elements by Type Ia SNe
to start at a higher \feh\ in the inner disk than in the outer regions.
More recent work by Gratton et al. (1996, 2000), Fuhrmann (1998, 2000),
Prochaska et al. (2000), Feltzing et al. (2003b), and Reddy et al. (2003)
suggests, however, that the differences are due to a chemical separation
between thin and thick disk stars. Thereby, these investigations
indicate that the thin and thick disks are discrete 
components, as originally suggested by Gilmore \& Reid (1983) from
a study of the distribution of stars in the direction of the
Galactic South Pole, and as also supported by kinematical studies
of unbiased samples of stars (e.g., Soubiran et al. 2003).
Further evidence of a chemical separation of thin and thick disk
stars is seen in \MnFe\ (Nissen et al. 2000; Prochaska \& McWilliam 2000),
in \EuBa\ (Mashonkina \& Gehren 2000, 2001; Mashonkina et al. 2003),
and perhaps in \ZnFe\ (Mishenina et al. 2002). Hence, the evidence
for a two-component (thin and thick disk) interpretation of the
\alphafeh , \MnFe, and \EuBa\ differences is quite compelling, although
one should note that some of the studies mentioned have selected stars
with extreme kinematics to make it possible to classify stars
as belonging to either the thin or the thick disk population.
If volume-limited samples of stars are selected, one may 
see more stars with intermediate abundance ratios, as 
suggested by the work of Edvardsson et al. (see Fig.\,\ref{nis.fig4}).  

As mentioned in the introduction, the survey of Edvardsson et al. (1993)
suggests that thick disk stars are older than thin disk stars.
Bernkopf et al. (2001) determined isochrone ages
for a few of the stars from Fuhrmann (1998, 2000) and obtained
ages between 12 and 14\,Gyr for the thick disk stars, whereas
the oldest thin disk stars have ages around 9\,Gyr. This points to
a hiatus in star formation between the thick and thin disk phases,
which nicely explains the abrupt decline in \alphafeh ; during
the hiatus, Type Ia SNe started to enrich the interstellar gas
with iron-peak elements, whereas the production of $\alpha$-capture
elements by Type II SNe stopped. Similarly, the decline in \EuBa\ is due to
enrichment of the interstellar gas with Ba from low-mass AGB stars,
while the high-mass stars had ceased to contribute Eu.

The work of Feltzing et al. (2003b) provides evidence
that \alphafeh\ in thick disk stars starts to decline
at a metallicity $\feh \geq -0.4$, and Mashonkina et al. (2003)
found a hint for a decline of \EuBa\ among thick disk stars at the
same metallicity. Hence, we may see a signature of the occurrence of the
products of Type Ia SNe  and low-mass AGB stars in the thick
disk at $\feh \simeq -0.4$, suggesting that the thick disk phase 
lasted at least $\sim$1\,Gyr.

As discussed in detail by Majewski (1993), there are two different
classes of models for the formation of the thick disk: the
pre-thin disk (top-down) models and the post-thin disk (bottom-up) models.
Within the first class, the chemodynamical model of Burkert, Truran, \& 
Hensler (1992) seems the most convincing. According to this model, the thick
disk is a stage in the collapse of the Galaxy where a high 
star formation rate leads to a high energy input into the interstellar medium,
halting the collapse and resulting in stars with a velocity dispersion
$\sigW \approx 40$\,\kmprs. As a result of metal enrichment, the cooling 
becomes more efficient, and the collapse continues forming the
thin disk from inside-out. A difficulty with this model is that
it predicts a thick disk phase with a duration of the order of 400\,Myr only, i.e. 
shorter than estimated above from the signature of enrichment by Type Ia SNe and
low-mass AGB stars in the thick disk. Also, it is not clear if the
model would agree with a 1--2\,Gyr hiatus in star formation between
the thick and the thin disk.

Among the class of post-thin disk models, violent heating of 
the early thin disk due to merging of a major satellite galaxy
(e.g., Quinn, Hernquist, \& Fullagar 1993) is the most obvious possibility.  
The thick disk stars were originally formed in the ancient thin
disk, with no tight limitations on the time scale for the chemical enrichment.
After the merger,
one can imagine that the star formation stopped for a while until the
gas assembled again in a thin disk, causing the hiatus that is needed
to explain the shift in \alphafeh\ and \EuBa\ between the thick and the
thin disk. Furthermore, if one assumes that the reestablished disk
is formed from gas in the thick disk plus accreted metal-poor gas from
the intergalactic medium, then one can explain why some thin disk stars
have a lower metallicity than the maximum metallicity of the thick disk,
i.e. the overlap of thin and thick disk stars in the metallicity range
$-0.8 < \feh < -0.3$.

As shown by Nissen \& Schuster (1997), there is also an overlap in metallicity
between halo and thick disk stars in the metallicity range
$-1.4 < \feh < -0.6$ with the majority of halo stars having lower
\alphafeh , \NaFe\ and \NiFe\ than thick disk stars. The low-\alphafeh\ stars 
tend to be on larger Galactic orbits than halo stars
having the same \alphafeh\ as thick disk stars. The explanation
may be that we have altogether two major components of our Galaxy: (1)
 a dissipative component consisting of the bulge, the {\it inner}
halo, and the thick disk all formed in a collapse stage with a fast star 
formation rate enabling the
metallicity to reach high values before Type Ia SNe and low-mass stars
started to enrich the gas, and (2) an accreted {\it outer} halo plus
the thin disk, where the star formation has proceeded on a longer time scale.
 
Much more work on stellar ages, kinematics, and abundances has 
to be carried out before we can be sure about the basic
scenario for the formation and evolution of our Galaxy. Many of the
results, quoted above, are based on studies of the chemical composition
of kinematically selected samples of stars. Thereby, the conclusions
may be affected by a kinematical bias, which, for example, exaggerates the
chemical separation between thick and thin disk stars. To avoid this,
it would be interesting to conduct an age-kinematics-abundance survey 
of, say, the 20 brightest stars in each of 25 metallicity groups spanning
the range $-2.0 < \feh < +0.5$ (i.e. a total of 500 stars), which should
be selected to lie on the main sequence and in the temperature range
$5000 \,{\rm K}\, < \teff < 6500$ K. It would also be very interesting to make
{\it in situ} studies of abundances and kinematics of stars in the
inner and outer halo, and at various places in the thin and thick disk.
Such studies may well give some surprises. Thus, Gilmore, Wyse, \& Norris (2002)
have recently conducted a low-resolution spectroscopic survey of $\sim 2000$
F--G stars situated 0.5--5\,kpc from the Galactic plane, and have found evidence
that the mean rotation velocity a few kpc away from the Galactic plane
is $\sim 100$\,\kmprs\ rather than the predicted $\sim 175$\,\kmprs\
from the local thick disk population.
Gilmore et al. propose that their outer sample is dominated by the debris stars
from the disrupted satellite that formed the thick disk. Clearly, it
would be very interesting to investigate the chemical composition of
such stars in detail.

\end{document}